\documentclass[10pt]{iopart}

\usepackage{graphicx}

\usepackage{hyperref} 

\newcommand{\rta}{\rightarrow}

\newcommand{\ep}{\epsilon}

\newcommand{\p}{\prime}
\newcommand{\om}{\omega}

\newcommand{\beq}{\begin{equation}}
\newcommand{\eeq}{\end{equation}}

\newcommand{\ben}{\begin{enumerate}}
\newcommand{\een}{\end{enumerate}}

\begin{document}

\title[Unconventional superconductivity in cuprates]{Leading theories of the cuprate superconductivity: a critique}

\author{Navinder Singh}

\address{Theoretical Physics Division, Physical Research Laboratory, Ahmedabad. India. PIN: 380009.}
\ead{navinder.phy@gmail.com; navinder@prl.res.in}
\vspace{10pt}
\begin{indented}
\item[]May 2020
\end{indented}

\begin{abstract}
We present a review of the leading theoretical approaches in the field of cuprate high temperature superconductivity. We start out by defining the problem and ask the question: whether an overarching theory possible (which is capable of explaining not only the mechanism of unconventional superconductivity but also a coherent understanding of the strange metal phase and the pseudogap phase)? If it is possible, what should we expect from the overarching theory? We list various experimental facts, and point out what can we learn from them and where do current theories stand in addressing them? Next, we present a critique of the current leading approaches. We conclude that although progress in the field has been unprecedented, but we still lack a coherent understanding. 
\end{abstract}

%
\vspace{2pc}

\vspace{2pc}

\noindent{``The first processes, therefore, in the effectual studies of the sciences, must
be ones of simplification and reduction of the results of previous investigations to a form in which the mind can grasp them." -- J. C. MAXWELL.}

\vspace{2pc}

%
\ioptwocol

\section{Introduction}
The problem of high temperature superconductivity in cuprates has been baffling people from last three decades. It is now understood that superconductivity in Cuprates (CupSCs)\cite{1}, in Heavy Fermion Superconductors (HFSCs)\cite{2}, and in Iron Based Superconductors (IBSCs)\cite{3} cannot be rationalized along the conventional Bardeen-Cooper-Schrieffer (BCS) paradigm of phonon mediated superconductivity. The experimental discovery of superconductivity in HFSCs is older than that in Cuprates, and IBSCs  are discovered in 2008\cite{3}. The mechanism of superconductivity in these systems is called unconventional (or non-phononic).

There is definite progress and there has been great many approaches\cite{4,5,6,7,8,9}. Some proved wrong and are out-dated\cite{4}. Some seems to be on the right track and have some elements of truth (as is discussed in subsequent sections)\cite{10}. But none of these is an overarching theory--many issues remain with them.

\section{Defining the problem}
The very first question that arises in the definition of the problem of unconventional superconductivity is that whether there is a single overarching framework in which the problem of superconductivity (SC) in CupSCs, IBSCs, and HFSCs can be rationalized, or, the idea of single overarching framework may not work in these apparently diverse systems?

The idea of a single theoretical framework seems reasonable from the point of view that all these systems show nearness to magnetic instability in their phase diagrams, superconductivity exhibits a dome-type shape in the tuning-parameter and temperature phase diagrams,  and all of them exhibit similar strange metal behaviour just above the superconducting dome in the normal state out of which superconductivity emerges on cooling\cite{taillefer1,scala1,stewart1}.

If such a framework exists then what should be the criterion that one can declare that the problem of unconventional superconductivity is considered resolved within that framework. The issue of a well define criterion or a metric to declare that the problem is considered solved and settled has been raised in\cite{kk}. Although we do not attempt to give a quantitative metric to resolve this issue, but in the following paragraph, we list important points which one should expect from a successful theory of unconventional superconductivity\footnote{More important requirements are discussed after we review main experimental facts in section (4).}:

\vspace{4pc}

\begin{itemize}

\item THE theory must have predictive power, that is, it should give us a value of $T_c$ (in a reasonable agreement with the experimental value) when other relevant parameters of a given material are put into the theory. In this way, it will address the question why does $T_c$ vary from one family to another (for example, it should tell us why YBCO superconduct at 90 K, and LSCO at 40 K).

\item THE theory must involve minimum amount of guess work, i.e., if there are a-priori assumptions, then these must be justified a-posteriori (via experimental confirmation and then mathematical proofs of the a-priori assumptions, if possible). In more colloquial language a theory based only on {\it watertight} arguments can pass this test. And it must satisfy the requirements of internal consistency.

\item THE theory must address a variety of experimental observations (not a selected set of observations suited for a given theoretical framework).

\item THE OVERARCHING THEORY should supply a transparent understanding of the Pseudogap (PG) phase in cuprates, and the strange metal behaviour in all the unconventional superconductors.

\item THE theory must give a semiquantitative account of perturbations that give rise to small changes in the value of $T_c$\cite{kk}.

\end{itemize}

\vspace{4pc}

These are the obvious requirements for any valid scientific theory. The problem is hard due to its strong coupling nature and lack of an appropriate mathematical framework\cite{5}. However, one can say that this is the most suitable time to attack this problem. Initial ``dust" has settled\cite{4}, and precise experimental information from a great variety of experimental probes is available\cite{stewart1}. And there is a significant progress in the theoretical understanding also\cite{5,10}.

In the current study we will focus only on the problem of unconventional superconductivity in the cuprates.

\section{Summary of our current understanding of the leading theoretical approaches}

In the following subsections we outline the main theoretical ideas emphasizing their physical basis rather than going into mathematical and technical details. To set a stage for discussions, schematic phase diagram is drawn, and the basic electronic structure is discussed, focusing on the active degrees of freedom that are responsible for unconventional superconductivity and magnetism.

There are five leading camps that attempt to address the mechanism of unconventional superconductivity in cuprates.\footnote{There are many other approaches which are not actively pursued at the current time and their criticism is summarized in (refer to\cite{4}).}

\vspace{4pc}

\ben

\item Anderson and collaborators (RVB/RMFT/Pain Vanilla).

\item Unconventional superconductivity from the exchange of magnetic spin fluctuations (an approach on the conventional BCS lines).

\item Unconventional superconductivity from repulsion.

\item The Kohn-Luttinger idea and its generalizations.

\item Cluster extensions of DMFT.

\een

\vspace{4pc}

Before we discuss these approaches one by one, let us review very briefly the typical phase diagram of cuprates and their electronic structure.
\begin{figure}[!h]
\begin{center}
\includegraphics[height=4cm]{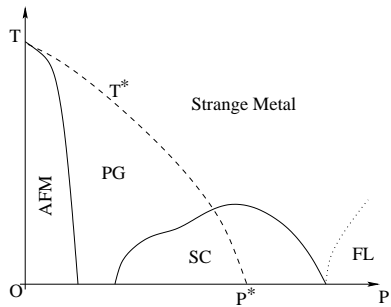}
\caption{A schematic cuprate phase diagram.}
\end{center}
\label{f1}
\end{figure}
Undoped Cuprates are magnetic-Mott insulators and exhibit AntiFerroMagnetic (AFM) order below a Neel temperature of the order of several hundred Kelvin (in $La_{2-x}Sr_x CuO_4$ or LSCO in short, it is  $T_N\simeq 300~K$). On hole doping (via substitution of $La$ with $Sr$ on a small fraction of $La$ sites) the AFM oder vanishes (figure 1). Similar trend is seen on the electron doped side\cite{greene}. Further hole doping leads to superconductivity (SC) in the form of a dome shape in the temperature-doping phase diagram. Dashed line marked with $T^*$ in figure (1) represents a cross-over to the pseudogapped regime (observed in many spectroscopic, thermodynamical, and transport probes)\cite{taillefer2}. For example, in ARPES (Angle Resolved Photoemission Spectroscopy) it is a partial gap observed in the spectral function (in the anti-nodal direction in the Brillouin zone). Just above the SC dome there is a metallic phase with anomalous transport and spectroscopic properties. It is called the strange metal phase (the most prominent and famous feature is its T-linear resistivity)\cite{taillefer2}. In the overdoped side, when superconductivity (SC) is no more stable, one has a good metal phase in the sense that resistivity scales as $T^2$ (a signature of a Fermi liquid).

\begin{figure}[!h]
\begin{center}
\includegraphics[height=4cm]{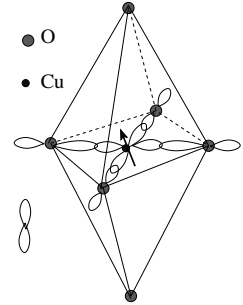}
\caption{The octahedron cage.}
\end{center}
\label{f2}
\end{figure}

Crystal structure of cuprates is lamellar in the form of $CuO_2$ layers sandwiched between spacer layers\cite{10}. From electrical resistivity per layer and from its anisotropy (along the ab-plane and along the c-axis direction) it is concluded that active elements where conduction happens are the $CuO_2$ layers\cite{4}. These are also the seat of magnetic degrees of freedom\cite{4,10,ander10}. Typical cuprates have an  octahedron cage (figure 2) which is elongated along the c-axis direction due to the Jahn-Teller distortion (it is very pointy in LSCO\cite{10}).

\begin{figure}[!h]
\begin{center}
\includegraphics[height=6cm]{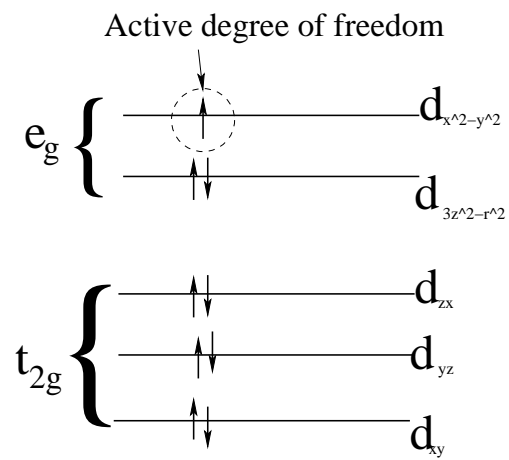}
\caption{Orbital structure and ground state occupations.}
\end{center}
\label{f3}
\end{figure}

The most relevant degrees of freedom in the electronic structure are depicted in figure (3)\cite{10}. d-orbitals of copper atom (in $d^9$ valence state in the undoped cuprates) are split into two sub-groups ($e_g$ and $t_{2g}$) due to the crystal field effects. $t_{2g}$ form the lower energy set and are filled first whereas $e_g$ form the higher energy set and filled at the last. Out of the nine electrons, six are filled in $d_{xy},~d_{yz}$, and $d_{zx}$, and the remaining three goes into $d_{x^2-y^2}$ and $d_{3z^2-r^2}$. Two electrons are filled in $d_{3z^2-r^2}$ and remaining one electron goes into $d_{x^2-y^2}$ orbital. The reason for this distribution is that in the orbital $d_{x^2-y^2}$ electrons face repulsion from {\it four} planner negatively charged oxygen atoms. In $d_{3z^2-r^2}$ orbital, electrons face repulsion from {\it only two} apical negatively charged oxygen atoms. Thus $d_{3z^2-r^2}$ orbital is in the lower energy state and must fill first and $d_{x^2-y^2}$ orbital is in the higher energy state, and it must fill at the last. Hybridization does not radically modify this picture\cite{10}. Thus, lone or un-paired electron remains in $d_{x^2-y^2}$ orbital, and it exhibit dual character: it is responsible for both magnetism (in the un-doped state) and it is also responsible for electrical conduction and superconductivity (in the doped state)\cite{10,ander10}. On doping this is the electron which is removed as it is the most ``unhappy" electron (\`a la Anderson)\cite{10}). But, even after 30 years, this issue remains contentious. There is a school of thought which believes that the hole doping removes electrons from oxygen p orbitals. And there are two sub-systems: itinerant and localized\cite{11}. We will not venture into this debate here, an interested reader can browse the references quoted. A summary is presented in Ref\cite{nav}.

With this very brief introduction to the phase digram and electronic structure, let us move on to the discussion of the leading theoretical approaches in understanding the mechanism of superconductivity in cuprates.

\vspace{2pc}
\noindent{\bf(i)\emph{Anderson and collaborators (RVB, RMFT, and Pain Vanilla).}}
\vspace{2pc}

One of the early approach is Anderson's RVB (Resonating Valence Bonds) idea. The physical picture of the original idea of the RVB state is of the following kind\cite{6}. Due to strong correlations the unpaired electrons in $d_{x^2-y^2}$ copper orbitals are localized in the undoped state. Spins of unpaired electrons form singlet pairs that resonate between near neighbours (This is much like Linus Pauling's idea of quantum mechanical resonance that gives increased stability to chemical bonds\cite{pauling}. For example in the benzene ring double bonds and single bonds resonate among themselves (i.e., there exists resonating Kekule structures)).

Anderson argued that this RVB state can be formally obtained from a Gutzwiller projected BCS pair superconducting state\cite{10}. {\it This is the fundamental assertion on which the entire RVB theory is crafted, and this has been the most controversial assertion}\cite{10}. The Gutzwiller projection is a way to enforce no double occupancy on Cu sites\cite{bas}. To execute it exactly is a great challenge. The slave-Boson formulations and the Renormalized Mean-Field Theory (RMFT) implement it approximately. More accurate approach is the variational Monte Carlo of the projected wavefunction\cite{6,param,lee}. The resulting phase diagram is depicted in figure (4). 

It has several attractive features. It is able to predict $d-$wave symmetry of the order parameter. The theory could qualitatively explain the dome type structure of the superconducting phase, which can be motivated in the following way. Below a temperature $T^*$ a ``spin gap" $\Delta$, which decreases on increasing doping, is predicted\cite{6,param}. This is due to the superexchange interaction mediated d-wave spin-singlet pair formation. Another temperature $T_{coh}$ below which phase coherence sets is also predicted\cite{lee}. This scale increases with increasing doping. Superconductivity is possible only below these two temperatures thus motivating the dome type structure (figure 4). Another way to motivate the dome type structure within the RVB scenario is to invoke the lack of the electronic compressibility in the extreme underdoped regime\cite{bas,bas1,bas2}. This is a great success of the theory, but it has many problems as discussed in the following paragraphs.

\begin{figure}[!h]
\begin{center}
\includegraphics[height=5cm]{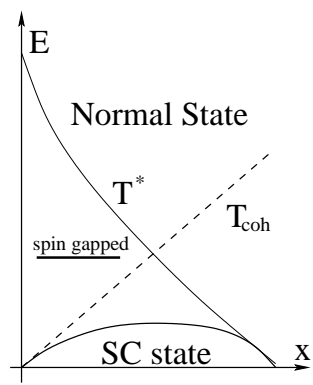}
\caption{The RVB plain vanilla phase diagram.}
\end{center}
\label{f4}
\end{figure}

The physical picture that emerges from the RVB and related theories is that the superexchange interaction $J$ is the cause of pairing. According to Anderson, pairing is instantaneous, not mediated by any ``glue" in the conventional sense (i.e., not a retarded interaction as in the BCS theory). In d-wave pairing state electrons simply avoid being very close to each other (due to central node) thus Coulomb repulsion is reduced\cite{6,10}.

The plain-vanilla RVB theory suffers from many problems. The basic premise on which it rests is that the ground state of cuprates is the Gutzwiller projected BCS state does not have a priori justification.\footnote{However, in condensed matter physics most of the time the ``reductionistic" approach is not possible, and progress is often achieved by proposing a reasonable ansatz for a possible ground state wavefunction, and then working out the consequences that can be compared with experiments.} It has only a posteriori justification in that it is able to reproduce the dome type structure. But if we identify ``spin gap" $\Delta$ of the theory with the pseudogap state, one runs into problems. The most prominent one is that it ends where the superconducting order parameter ends (figure 4) as if pseudogap phase is the ``mother phase" of superconductivity (SC). But it is in direct contradiction to many experiments\cite{varma} (such as low temperature electronic heat capacity measurements, NMR, Hall effect measurements, Nernst effect measurements etc\cite{taillefer2}) in which pseudogap phase ends somewhere in the middle of the superconducting dome (figure 1) at a doping $p^*$. Several experiments show that $p^*$ is an AFM Quantum Critical Point (QCP).  At $p^*$ several anomalies exist, like electronic heat capacity shows logarithmic divergence ($C_{el} \propto -T \ln(T)$), resistivity is perfectly T-linear, and from Hall coefficient, carrier density shows a jump from $1+p$ holes per copper atoms to $p$ holes per copper atom as the doping is reduced through $p^*$\cite{taillefer2}. As discussed in sections 4 and 5, anomalies at $p^*$ are very important incontrovertible facts, and these must form an integral part of a successful theory. But the RVB plain-vanilla is continuous across $p^*$\cite{varma}. This is the major weakness of the theory. It fails to capture this. In addition, the strange metal phase remains beyond its scope. 

Thus, we can say that the RVB plain-vanilla is a partly successful theory. Whether it can be extended to include the above mentioned important points remains an open issue.

\vspace{2pc}
\noindent{\bf (ii)\emph{Unconventional superconductivity from the exchange of magnetic spin fluctuations (an approach on the conventional BCS lines).}}
\vspace{2pc}

Along the traditional BCS lines, spin-fluctuation glue theories has been developed early on by Pines, Scalapino, Ueda, Moriya, and their collaborators\cite{moriya2,moriya1,loh,scala10,scala11,scala1,millis,moth1,moth2,moth3}. It is argued that electrons are paired up by the exchange of magnetic spin fluctuations (just like in the BCS theory where phonons provide the glue). A clear implementation of this approach requires two-types of electronic sub-systems in $CuO_2$ planes\cite{11}:

\begin{itemize}
\item Itinerant carriers which form Cooper pairs.
\item{Localized spins that provide the ``glue" in the form of magnetic spin fluctuations}.
\end{itemize}

It is argued that doping leads to holes in oxygen $p$ orbitals and these constitute an itinerant sub-system. Copper $3d_{x^2-y^2}$ orbitals contain a spin degree of freedom and it constitutes a localized sub-system. Due to doping and hole motion, AFM order of the Mott insulating regime ``melts".   Below the  transition temperature $T_c$ mobile carriers are paired up by the magnetic spin fluctuations in the localized spin system. Spin fluctuations in the localized sub-system provide the required ``glue". 

In addition, scattering of the mobile carriers off the spin fluctuations in the localized sub-system leads to anomalous transport, thermodynamical, and spectroscopic properties (i.e., the strange metal behaviour). 

This approach is conceptually very appealing but it suffers from several problems. The division of electrons in $CuO_2$ planes into two sub-systems is rather controversial. Anderson argued that the relevant electrons in $Cu~3d_{x^2-y^2}-O~2p$ hybrid orbitals has dual character\cite{ander10}. The very same carriers which conduct and superconduct (depending on temperature) at finite doping becomes localized magnetic degrees of freedom at zero doping. However, the two-component approach advocates a different view\cite{11}: In the two-component approach doped holes are thought to exist on oxygen p-orbitals and constitute mobile degrees-of-freedom, and unpaired electrons on copper sites provide the magnetic glue.

On the contrary, if we assume that it is a single system and still use the pairing glue idea, then it is difficult to imagine that the same electrons being paired among each other and the very same electrons also providing the glue\cite{5}. Thus, we face conceptual problems.

Putting these conceptual difficulties aside let us see how these ideas are mathematically implemented and executed. Moriya etal uses dynamical magnetic susceptibility $\chi(q,\om)$ computed by using the Self-Consistent Renormalization (SCR) theory in the modified gap equation as an effective interaction and it also results in the d-wave gap symmetry (pairing potential $\propto Im\chi(q,\om)$)\cite{moriya1}. The SCR goes beyond the Random Phase Approximation (RPA) in that it takes into account the effect of thermally excited magnetic fluctuations on the ground state in a self consistent manner\cite{moriya10}. $\chi(q,\om)$ can also be computed beyond RPA using the FLEX approximation\cite{bickers}, and using the Vilk-Tremblay approach\cite{vilk}.

As stressed, these mathematical formalisms suffer from transparency issues: which electrons are being paired and which electrons provide the ``glue"?  Also consider the process of condensation in a little more detail. When electrons are being paired and are condensing into the condensate, {\it the ``glue" $\chi(q,\om)$ will itself be weakening.} What is the basis that the ``glue" itself survive the condensation transition (if the single band Hubbard model is used)? In some of the studies, experimentally determined $\chi(q,\om)$ is used. These studies are, therefore, called phenomenological (not microscopic). For a review refer to\cite{scala10}. But the same criticism can be raised to these studies also. 

On the other hand, Ueda-Moriya's SCR theory has been quite successful in addressing the magnetism of weakly and nearly ferromagnetic transition metals (it even provided the physical basis for the Curie-Weiss law in an itinerant system!)\cite{moriya10}. It has even addressed the problem of T-linear resistivity in 2D AFM itinerant metals, and they argue that it can be taken as a model for the strange metal regime in cuprates\cite{moriya1}. However, more physical questions remain to be addressed, like, which electrons scatter off which electrons, and how the momentum is being degraded (electron-electron interactions in a single band conserve momentum). 

Thus, if the mechanism of momentum degradation becomes clear in the SCR theory, it can provide a model for the strange metal regime. But it cannot be a valid case of the unconventional superconductivity as it suffers from the above mentioned problems.

Another serious criticism has been raised by Phil Anderson when Migdal-Eliashberg theory is used to study the strong coupling problem in this magnetic glue scenario\cite{ander20}. He argues that the Eliashberg extension of the BCS theory describes electron pairs bound together by an exchange of low-frequency bosons.   { \it However, in cuprates exchange of AFM spin fluctuations is a very high energy (high frequency) phenomenon. Thus the Eliashberg theory cannot be applied to the cuprate problem. Both the scales $J$ and $U$ constitute a very high energy (high frequency) dynamics.} Thus it is logically not justified to apply the Eliashberg theory when a slow degree of freedom cannot be rigorously proved from the microscopic considerations.

Therefore, the formulation of the cuprate problem along the ``glue" ideas suffers both from conceptual and from concrete difficulties.

\vspace{2pc}
\noindent{\bf(iii)\emph{Unconventional superconductivity from repulsion.}}
\vspace{2pc}

In this scenario, unconventional superconductivity is argued to be possible even in a purely repulsive interactions (all repulsive scenario)\cite{raghu}. It can be intuitively motivated in the special case of weak coupling of the Hubbard model in which the Hubbard $U$ is much less than the nearest neighbour hopping energy $t$ ($U<<t$). The gap structure is obtained from the solution of a BCS type equation:

\beq
\Delta_k = -\sum_{k^\p} \Gamma_{k,k^\p} \frac{\Delta_{k^\p}}{2 \sqrt{\ep_{k^\p}^2 +\Delta_{k^\p}^2}}
\eeq

Here $\Gamma_{k,k^\p}$ is an effective repulsive interaction (a renormalized two-particle vertex function)\cite{raghu}. It is argued that if $\Gamma$ is sufficiently $k$ dependent function and peaks at a special wave vector $Q$, then a sign changing gap ($\Delta_{k+Q} = -\Delta_k$) is a solution of the above gap equation. This again leads to the correct d-wave gap function!

However, this is only in the weak coupling limit. In the intermediate coupling regime ($U\sim t$), which is relevant for cuprates, no such conceptually transparent picture is available. It turns out that the wave vector $Q$ is the AFM wave vector of the AFM phase in the undoped system. On doping, AFM correlation survive even upto the optimal doing regime\cite{5}. Thus the problem of strong correlation remains in the major part of the phase diagram. In a simple language, electron motion is highly correlated via magnetic fluctuations even at optimal doping\cite{5}.

This theory successfully yields d-wave nature of pairing. However, the prediction of the dome type structure of the superconducting order parameter as a function of doping and anomalies at $p^*$ remain outside the scope of this idea (additional ideas like ``The Emery-Kivelson argument" are required to qualitatively explain the dome structure\cite{kivv}). Another way to get the dome is to investigate the doping dependence of $\Gamma_{k,k^\p}$. A microscopic model for {\it doping dependent} two-particle vertex function needs to be worked out. The pseudogap and the strange metal phase also remains out of the scope of this approach.


\vspace{2pc}
\noindent{\bf(iv)\emph{The Kohn-Luttinger idea and its generalizations.}}
\vspace{2pc}

In this school of thought, it is argued that electrons are paired up by the attractive part of the otherwise repulsive electron-electron interactions. The basic idea behind this is the Friedel oscillations. There are attractive components of screened $V(r)$ at large $r$ (figure 5). Translating it into momentum space, it turns out that the interaction has attractive components for large angular momentum quantum numbers ($m$).

\begin{figure}[!h]
\begin{center}
\includegraphics[height=3.5cm]{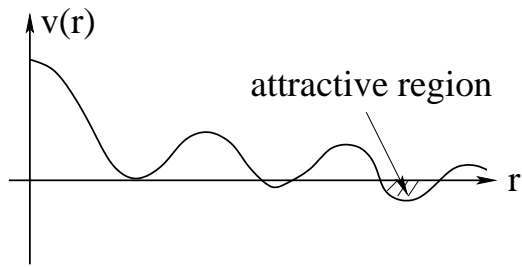}
\caption{A cartoon diagram showing attractive parts of the screened electron-electron interaction.}
\end{center}
\label{f5}
\end{figure}

More precisely, the Kohn-Luttinger idea can be stated in the following way. Fully screened Coulomb repulsion, in isotropic systems, can have attractive components for large odd $m$\cite{kohn}. It turns out that the pairing problem decouples for different values of $m$ in an isotropic system\cite{lif}. Thus superconductivity, in principle, is possible even if one $m$ provides an attractive channel.  This is another example of  ``superconductivity from repulsion".

In 1968, Fay and Layzer\cite{fay} proved that for the Hubbard model with small repulsive $U$, attraction emerges in the {\it 2nd order} in $U$ for all channels $m\ge 1$ (it is maximum for $m=1$, that is, in the $p-$wave channel).

The Kohn-Luttinger and the Fay-Layzer analysis cannot directly be applied to lattice systems like for cuprates. However, Chubukov and collaborators have generalized this analysis to lattice systems, and applied to cuprates\cite{maiti}.  They discover $d-$wave superconductivity. The  basic idea can be expressed in the following way. Look at the regimes of the Brillouin zone where the Electron Density of States (EDOS) is very large. This occurs in the anti-nodal directions in the Brillouin zone (figure 6). 
\begin{figure}[!h]
\begin{center}
\includegraphics[height=3cm]{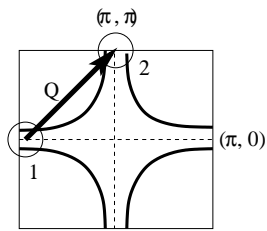}
\caption{AFM wave vector connecting two regions with large EDOS.}
\end{center}
\label{f6}
\end{figure}
The vector connecting these two patches (patch 1 and patch 2) is roughly the AFM wave vector $Q$. Interactions are all repulsive. The question is how can one get an attractive interaction? Following Chubukov, let us denote intra-patch (within a patch) repulsive interactions by $g_1$ and inter-patch (between the patches) repulsive interactions by $g_2$. Consider two effective coupling constants: $\lambda_a = g_1+g_2$, and $\lambda_b = g_1-g_2$. Perform Kohn-Luttinger analysis for on-site repulsion $U$. To first order in $U$ no $\lambda$ is negative (remember for paring we need at least one lambda negative). But, as in the case of Fay and Layzer, to second order in $U$, $\lambda_b$ turns out to be negative! And superconductivity becomes possible, but in this case in the $d-$wave channel, in which the order parameter changes sign from patch 1 to patch 2. Thus, this approach too leads to $d-$wave pairing!

However, the Kohn-Luttinger (KL) idea is for dense electronic systems where occasionally over-screening may happen. But cuprates are low electronic density systems thus screening is significantly less. KL idea may be applicable on the overdoped side of the phase diagram where large Fermi surface is observed and $2~k_F$ related Friedel oscillations can be present. However, in the underdoped regime, it is the Mott physics and strong electronic correlations which dominate the physics. Thus applicability of the Kohn-Luttinger idea to cuprates is not beyond criticism. In addition, it misses out the important physics at $p^*$ which must be an integral part of any successful theory (as discussed in sections 4 and 5).


\vspace{2pc}
\noindent{\bf(iv)\emph{The cluster extensions of DMFT.}}
\vspace{2pc}

The cuprate problem is an essentially non-perturbative in nature when formulated in terms of the simple Hubbard model. Out of the various numerical approaches, the cluster extensions of the Dynamical Mean Field Theory (cDMFT) has been particularly successful in reproducing some parts of the phase diagram, including the pseudogapped state\cite{sordi1,maier,gull1,gull2, simard,fratino,sordi2,sordi3,alloul1,shiro01,shiro02}. However, several issues remain within these approaches as well.

Cellular or cluster DMFT fully takes into account the short-range correlations within the cluster. In the calculation of the superconducting state, the gap equation is not used. The anomalous self-energy is simply assumed to be finite. If it becomes finite when the impurity problem is solved self-consistently, then it produces an anomalous component in the dynamical mean-field, and thus a self-consistent superconducting solution is obtained\cite{shiro1}. Because of the self-consistency condition, it is a more rigorous approach as compared to an approach based on a-priori superconductivity ansatz.

The cDMFT based approaches do lead to a phase diagram which qualitatively resembles the experimental phase diagram\cite{sordi1,maier,gull1}. The cDMFT calculations performed by Sordi et al\cite{sordi1} show that the PG phase is disrupted by the occurrence of the d-wave superconductivity dome. The authors show that the PG line is controlled by the so-called ``Widom line", which is a locus of the maxima of the thermodynamical functions in the co-existence regime (Widom line is an extension of the first order phase boundary beyond the critical point into the co-existence regime\footnote{To learn more about the Widom line in the context of classical phase transitions, refer to: Limei Xu et al, PNAS November 15, (2005) 102 (46) 16558-16562; https://doi.org/10.1073/pnas.0507870102.}). The $T^*$ line defined by the maxima in the static susceptibility is a precursor of the Widom line (refer to figure 3 in\cite{sordi3})  and it meets the SC dome near optimal doping\cite{sordi1,sordi2,sordi3}. The main point is that the phase diagram obtained using cDMFT is closer to the experimental phase diagram as compared to that obtained via other approaches (especially in regard to the $T^*$ line).

Similar phase diagram (in which the $T^*$ line meets near the top of the SC dome) is also obtained in the DCA calculations of Maier and Scalapino\cite{maier} and in the calculations of Gull and Millis\cite{gull1,gull2}.

Although these approaches seem to be much more successful in obtaining the correct $T^*$ line, but the strange metal behaviour and its connection to the unconventional superconductivity remains an open issue (i.e., the connection between T-linear resistivity and the pairing correlations). However, there is some success in this direction. It turns out that the maximum in the single-particle scattering rate for c-axis transport is near the maximum $T_c$ (blue diamonds in figure 2 in\cite{fratino}). This could be a beginning of the numerical deduction of the very important relation: $A \propto T_c$.\footnote{Computing the in-plane resistivity has not been done because of the difficulties associated with vertex corrections. These vertex corrections can be neglected in the direction perpendicular to the plane. Private communication with A.-M.S. Tremblay.} 

Another important question is related to the QCP at $p^*$.  As discussed before the behaviour of heat capacity and the Hall coefficient is very anomalous at this point. It seems that cDMFT approaches has partial answers to these questions also.\footnote{There is a classical critical point and it is connected to $T=0$ line with a single first order phase transition with hysteresis. At large enough $U$, the critical point is at so low a temperature that it may look like a QCP. In fact, it is found that the heat capacity does show the expected  $-T\ln T$ behaviour (refer to Fig. 2 in ref\cite{tr1}, and also refer to\cite{tr2} in which the observed maximum in the heat capacity near the critical point is reproduced). Private communication with A.-M.S. Tremblay.}.

Given all the pros and cons of cDMFT based approaches,  but one point is clear.  The reason for pairing in these approaches is not made explicit in the sense that the conceptual picture like that of, for example,  the Kohn-Luttinger theory (attractive components of the otherwise repulsive Coulomb interaction) or like that of paring by a bosonic glue is missing in cDMFT based approaches. Whether this question should be the part of ``the theoretical minimum" of the high-$T_c$ problem is itself a debatable issue.

\section{Some selected known experimental facts, and what can we learn from them?}

\begin{enumerate}

\item It is observed that $ab-$plane DC resistivity ($\rho_{ab}$) per $CuO_2$ plane for a variety of cuprates is roughly the same\cite{bat,4,bari}. This observation clearly shows that charge carriers reside in $CuO_2$ planes and scattering also happens inside the planes itself. This establishes the two dimensionality of the cuprate problem\cite{4}.

\item Electronic structure considerations and Mott physics at zero doping show that dominant interaction among un-paired electrons is repulsive in nature. And its effect remains important even upto the optimal doping\cite{5}. All other interactions (like phonons) are sub-dominant\cite{10}.

\item Experimentally (via Wiedemann-Franz law) it is observed in $Nd-LSCO$ that ground state of the pseudogapped (PG) phase just below $p^*$ (when SC is suppressed using high magnetic fields) is indeed a good Fermi Liquid (obey Wiedemann-Franz (WF) law just as simple metals do)\cite{taillefer2}.\footnote{Any proposed theory of the PG state must respect this fact.} Thus the ground state is not exotic. Only at $p^*$, where PG ends, some observables show anomalous behaviour (like, electronic heat capacity shows a logarithmic divergence, resistivity becomes perfectly $T-$linear etc)\cite{taillefer2}. Verification of the WF law at very low temperatures in other cuprates is much desired (although experimentally very challenging).

\item At optimal doping DC resistivity is perfectly $T-$linear over many decades of temperature. Away from optimal doping but above the superconducting dome there is a $T-$linear component and a $T^2-$component of resistivity ($\rho = A T +B T^2$). {\it But important point is that as far as the system is superconducting there is a $T-$linear component above the dome.} And the coefficient of the $T-$linear component is proportional to the value of $T_c~:~A \propto T_c$. {\it It means that the mechanism responsible for $T-$linear resistivity is connected in an important way to the mechanism of superconductivity (SC)}\cite{taillefer1,taillefer2}.

\item In the strange metal regime above the dome, the Angle Dependent Magneto Resistance (ADMR) data has been modeled with two scattering rates: (1) a $T^2$ dependent rate which is isotropic around the Fermi surface, and (2) an anisotropic $T-$linear scattering rate which is not uniform around the Fermi surface. This scattering rate is maximal along the anti-nodal directions of the Brillouin zone\cite{taillefer2,abdel}. This anisotropy of the scattering rate is directly connected with $d-$wave nature of the pairing correlations. {\it This further support the fact that the mechanism of $T-$linear resistivity and the mechanism of $d-$wave SC are intimately connected to each other.}

\item The above point regarding the connection between $T-$linear resistivity and superconductivity is further highlighted: It is experimentally found in LSCO\cite{cooper} that when superconductivity is suppressed by a sufficiently high magnetic field, $T-$linear regime of DC resistivity is found to fan out (figure 7). This points towards the fact the pairing in cuprates has much to do with the mechanism which leads to $\rho\propto T$.

\begin{figure}[!h]
\begin{center}
\includegraphics[height=3.5cm]{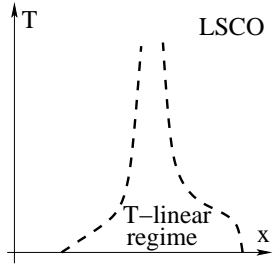}
\caption{$T-$linear regime of DC resistivity fans out in the shape of a dome when superconductivity is suppressed by high magnetic field.}
\end{center}
\end{figure}

\item AC analogue of $T-$linear resistivity is $\om$-linear generalized Drude scattering rate\cite{4}. In most of the systems it is found that $\frac{1}{\tau(\om)} \propto \om$. It is another important feature of the strange metal regime (figure 1). Understanding of $T-$linear DC resistivity can lead to its understanding\footnote{Private communication with Phil Anderson.}.

\item Hall angle $\Theta_H$ from the Drude theory is given by $\tan\Theta_H =\frac{\sigma_{xy}}{\sigma_{xx}} = \om_c \tau_H$ where $\om_c$ is the cyclotron frequency, $\sigma_{xy}$, and $\sigma_{xx}$ are the transverse and longitudinal conductivities, respectively. Experimentally, it is found that $\frac{1}{\tau_H} \propto T^2$\cite{4,col}. But for electrical resistivity $\frac{1}{\tau} \propto T$. Thus, two scattering rates scale very differently with temperature.  How to reconcile these observations (or ``two-lifetime picture") with the other features of the strange metal regime? This remains an outstanding open problem\cite{col}. 

\begin{figure}[!h]
\begin{center}
\includegraphics[height=4cm]{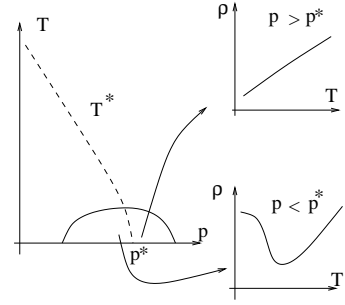}
\caption{Metal-to-Metal transition: Resistivity as a function of temperature shows an up-turn below doping $p^*$. It indicates loss of charge carriers below $p^*$ (including inputs from the Hall effect measurements).}
\end{center}
\end{figure}

\item Metal-to-metal transition: At $T\rta 0$, the PseudoGap Quantum Critical Point (PG QCP) is at $p^*$\cite{taillefer2}. When the doping is $p > p^*$ resistivity drops roughly linearly with decreasing temperature. But at $p<p^*$ there is a pronounced upturn in DC resistivity (figure 8), and at very low temperatures it tends to saturate. This is a metal-to-metal transition (as stressed by Louis Taillefer). The rise in resistivity is attributed to loss of charge carriers when systems enters into the PG phase (this is inferred from the Hall coefficient measurements also)\cite{taillefer2}. The carrier number reduces from $1+p$ holes per $Cu$ atom at $p>p^*$ to $p$ holes per $Cu$ atom at $p<p^*$. {\it But why does carrier number drop below $p^*?$ Answer to this question hides the secret of the PG phase. This observation is very important.}

\item Issue of phonons: At optimal doping there is no isotope effect and DC resistivity is $T-$linear from very low temperatures (milli Kelvin) to very high temperatures (about thousand Kelvin)\cite{taillefer1,cooper}. It does not show $T^5$ bending at low temperatures(figure 9). This is a proof that phonons are not the main players. It might be coming from some sort of magnetic scattering. However, at underdoping there is an isotope effect\cite{361,362,363}. Does it mean that at underdoping superconductivity is phonon mediated? Or, is it still electronically mediated but isotope effect is due to magneto-elastic effects? These questions need to be resolved. Or, phonons participate in a non-trivial way?

\begin{figure}[!h]
\begin{center}
\includegraphics[height=3cm]{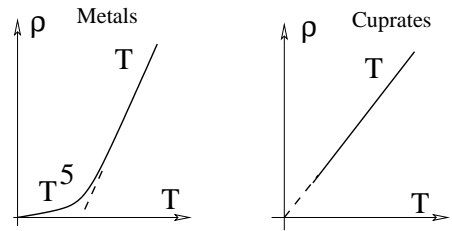}
\caption{DC resistivity (impurity subtracted) in a simple metal (with $T^5$ bending) and in cuprates ($T-$linear). Absence of $T^5$ bending rules out the importance of phonons in the mechanism of $T-$linear resistivity.}
\end{center}
\end{figure}

\item Cooper pair size and destruction of superconductivity in the underdoped regime: The size of Cooper pairs in cuprates is somewhere in the range of $10~\AA$ to $30~\AA$\cite{dago}. This is much smaller than that found in the BCS superconductors (by over an order of magnitude). This clearly shows that pairing in cuprates is {\it local} in nature.

There is a viewpoint known as {\it the Emery-Kivelson viewpoint} regarding superconductivity in the underdoped regime connected with the above fact\cite{emery}. There are two energy scales: (1) energy required to break a Cooper pair ($\Delta$), and (2) energy required to take a Cooper pair out of the condensate ($\rho_s$, in other words, it is the energy lowering coming from the collective phase coherence).    

It is argued that in the underdoped cuprates $\Delta$ is much greater than $\rho_s$. Thus, superconductivity in underdoped cuprates is destroyed not due to breaking of Cooper pairs, but it is destroyed due to the thermal destruction of the phase coherence (low $\rho_s$). This is just opposite to what happens in the conventional BCS superconductors where destruction of superconductivity starts with the breaking of Cooper pairs (as $\Delta<<\rho_s$ in the conventional BCS superconductors).

Indeed in the initial Nernst effect measurements SC fluctuations are found to be present in a large portion (inside the PG phase boundary) of the underdoped side of the phase diagram (upto $\sim 5~T_c$)\cite{wang,ong1,xu,ong2}. In this ``Princeton interpretation" phase fluctuations are argued to be the cause of SC destruction in the underdoped regime. However, recent refined Nernst effect measurements show that the fluctuation regime is very narrow ($\sim1.5~T_c$) and follow a narrow strip along the contour of the SC dome\cite{cyr}.

\item Non-magnetic impurities in $CuO_2$ planes: superconductivity in cuprates is very sensitive to non-magnetic impurities in $CuO_2$ planes. For example, when small fraction of the $Cu$ sites in YBCO is replaced by $Zn$ ($YBa_2Cu_{2.9}Zn_{0.1}O_{7-\delta}$) the value of $T_c$ reduces from $90~K$ to $50~K$\cite{stewart1}. This can be qualitatively addressed in the two famous approaches (i) and (ii) in section 3. In the Mott physics scenario hole doping destroys AFM order and system starts conducting (opening the quantum traffic jam). If some of the $Cu$ sites are replaced by $Zn$, then {\it local} quantum jamming effect occurs and {\it local} islands of AFM order are formed. This reverses the effect of hole doping thus suppress $T_c$. In little more detail, zinc doping leads to a closed d shell $d^{10}$ configuration instead of open $d^9$ configuration of copper. This leads to hindrance of hole motion around $Zn$ site, and local AFM arrangement on nearby copper sites re-appears. Thus, it is like re-doing the quantum traffic jam (which was lifted by hole doping) and $T_c$ is suppressed on $Zn$ doping\cite{fn1}. In the spin-fluctuation mediated pairing scenario $Zn$ doping leads to a weaker magnetic sub-lattice (as zinc creates non-magnetic sites). This is like weakening of the magnetic glue, and thus suppression of $T_c$. This is only a qualitative explanation. Also it needs to be seen that whether the RVB approach adopted in\cite{mai} leads to local AFM islands around $Zn$ sites? Only then full understanding of it can be achieved.

\item $T^2-$resistivity in electron doped cuprates: From $T_c$ upto $400~K$ in electron doped cuprates it is observed that resistivity is proportional to $T^2$. Naively, one can say that it comes from Fermi liquid physics. However, it may be due to strong electron correlations (not due to Fermi liquid physics)\cite{greene}. Resistivity behaves very anomalously in electron doped side also. Scaling behaviour of resistivity $\rho$ can be divided into three regions:

\begin{eqnarray}
&\propto& T, ~~~~~~~~~~~~mili~Kelvin<T<T_c  \nonumber\\
&\propto& T^2, ~~~~~~~~~~~T_c <T<400~K \nonumber\\
&\propto&  T^2 ~to~ T,~~~~400~K<T<1000~K    
\end{eqnarray}

This behaviour remains outside the scope of a comprehensive understanding. 

With the above list of some main experimental facts we can now ask the question: how well the theories address these experimental facts?

\end{enumerate}

\section{Where do the theories stand in addressing the above facts?}

\begin{itemize}

\item {\it None of the above theories microscopically establish the extremely important connection between mechanism of $T-$linear resistivity and the mechanism of superconductivity, more precisely the theoretical deduction of $A\propto T_c$ remains an open challenging problem outside the scope of all the proposed theories}\cite{fn2}.  From experimental side the connection is very clearly visible and has been stressed\cite{taillefer1}. This connection has been further supported by ADMR data (point (v) in the previous section).

\item At $p^*$ (the PG QCP) several observables show anomalous behaviour ($C_{el}\propto - T\ln T$, and $\rho\propto T$ etc). These are the signatures of an AFM QCP\cite{taillefer2}. Current theories are not in a position to reconcile these very important facts within their proposed schemes. cDMFT based approaches reproduce this logarithmic divergence in the heat capacity from a critical point inside the dome\cite{tr1,tr2}. However, the critical point is not exactly situated at the $T=0$ line. On the other hand, if one argues that this QCP at $p^*$ is a source of quantum critical fluctuations that binds the electrons in Cooper pairs, then one runs into problems as discussed in section 3(ii).

\item A successful theory must explain fanning out of the $T-$linear regime (figure 7), that is, when superconductivity is suppressed whatever metallic state one is left with, it should exhibit anomalous scattering in the form of $T-$linear scattering rate (author also suspect that when superconductivity is suppressed, the $\om-$linear generalized Drude scattering rate ($\frac{1}{\tau(\om)}\propto \om$) regime should also fan out). None of the theories is in a position to address this. The $\om-$linear generalized Drude scattering rate ($\frac{1}{\tau(\om)}\propto \om$) remains outside the scope of the above mentioned approaches. Only in the Moriya-Ueda SCR theory it is captured\cite{moriya1}. But it suffers from issues mentioned in 3(ii) when it is applied to the superconducting state.

\item Two life-time picture ($\frac{1}{\tau_H} \propto T^2$ and $\frac{1}{\tau} \propto T$) has also resisted a universally accepted view\cite{col}.

\item None of the above theories account for the sudden carrier number reduction when doping is reduced through $p^*$. Understanding of this fact is very important for the structure of any potential theory of the PG state. Also it is important in understanding the actual mechanism of SC, as at $p^*$, $T_c$ is maximum (optimal doping). It cannot be a coincidence. It is related to the very mechanism of SC.  Any successful theory of SC state must respect this fact.

\end{itemize}

Thus we observe that the solution of the problem of unconventional SC in cuprates is quite far from complete.

\section{A critique of current theories: summary points}

\vspace{3pc}

\begin{itemize}

\item In the plain-vanilla RVB phase diagram (figure 4) spin gap ends where superconducting order parameter ends (towards the overdoped side of the SC dome). But it is in direct contradiction to many experiments (like NMR, very low temperature heat capacity measurements etc). These experiments show that spin gap or the pseudogap ends somewhere in the middle of the SC dome (figure 1) at a critical doping $p^*$ which is a QCP showing logarithmic divergence in low temperature heat capacity ($c_{el} \propto -T\ln T$) and a perfectly T-linear resistivity. The plain-vanilla RVB is continuous across $p^*$. Thus it misses out the important physics at $p^*$. Another essential feature of $p^*$ is that for $p>p^*$ carrier density is $1+p$. It jumps to $p$ when $p$ is reduced through $p^*$. This is a very important feature and a successful theory of high-$T_c$ must be based on this. RV plain-vanilla misses this out also.

\item Theory of cuprates based on the ideas of BCS or Eliashberg extension of it (but replacing phonons with spin fluctuations) also suffers from conceptual (electrons pairing among themselves and the very same electrons providing the pairing glue) and factual issues (Anderson's criticism: $U$ and $J$ leads to a very high energy (high frequency) dynamics and the Eliashberg theory has no logical justification when applied to cuprates).

\item SC from repulsion is easy to grasp in the weak coupling limit ($U<<t$) where BCS type equation can be written. But in the intermediate coupling regime ($t\sim U$, actually relevant to cuprates) no such convincing picture available. As stressed before, the important physics at $p^*$ must be a part of a successful theory. It is not an epi-phenomenon. At this point $T_c$ is maximum (optimal doping) and this trend is found in all the unconventional superconductors (CupSCs, IBSCs, and HFSCs). It cannot be a coincidence.  All repulsive scenario fails to respect this fact.

\item Kohn-Luttinger idea may be applicable on the overdoped side of the phase diagram where large Fermi surface is observed and $2~k_F$ related Friedel oscillations can be present. However, in the underdoped regime, it is the Mott physics and the strong electronic correlations which dominate the physics. And there is a complete reconstruction of the Fermi surface when doping is reduced through $p^*$. How can KL formalism remains smoothly applicable when a drastic Fermi surface reconstruction happens at $p^*$? Thus it seems to  miss out the important physics at $p^*$ which must be an integral part of any successful theory.

\item The cluster extensions of the Dynamical Mean Field Theory (cDMFT) has been particularly successful in dealing with the strong coupling nature of the problem, and qualitatively able to reproduce the phase diagram. The cDMFT is a more rigorous approach as compared to that based on an apriori superconductivity ansatz, because the {\it self-consistency condition} is used to obtain a superconducting solution. Although these approaches seem to be much more successful in obtaining the correct $T^*$ line (meeting the SC dome at its top), but the strange metal behaviour and its connection to the unconventional superconductivity remains as an open issue. However, some success has been achieved in obtaining anomalous behaviour of heat capacity near the critical point.

\end{itemize}

\vspace{3pc}

In conclusion, the current paper serves two purposes: (1) It summarizes the leading theories of cuprate superconductivity and the main experimental facts, and (2) it presents a critique of the leading theories. From the analysis of the experimental facts we notice that the current popular theories of SC in cuprates suffer from several issues, both of conceptual and of concrete nature.

\vspace{3pc}



\section*{Acknowledgments}
Author would like to thank Hai-Hu Wen and Kishore Dutta for carefully reading the manuscript and suggesting corrections. He is thankful to Shiro Sakai and A.-M. S. Tremblay for their help regarding cDMFT approaches, and for suggesting references. Thanks are also due to G. Baskaran for suggesting some references, and to Lev Mazov for discussions. Author dedicates this manuscript to the memory of Prof. N. Kumar.

\section*{References}




\begin{thebibliography}{99}
\bibitem{fn1} In the Mott picture zinc doping is modeled as a disorder term in the Hubbard model and its solution using quantum monte carlo within dynamic cluster approximation leads to suppression of $T_c$\cite{mai}.
\bibitem{fn2} In the Moriya-Ueda SCR theory it is found that $T_c$ is proportional to $T_0$ which is a measure of the frequency spread of the wave vector dependent part of the spin fluctuations. Whether $T_0$ is proportional to $A$ (the coefficient of $T-$linear resistivity) for a set of doppings across the dome remains an open issue. That is, doping dependent verification of $A(x)\propto T_c(x)$ will be a real test.
\end{thebibliography}
\end{document}